# Computational Methods for Single-Cell Multi-Omics Integration and Alignment

Stefan Stanojevic[2], Yijun Li[1], Lana X. Garmire[1,2] *

[1]*Department of Biostatistics, University of Michigan, Ann Arbor, MI, USA*

[2]*Department of Computational Medicine and Bioinformatics, University of Michigan, Ann Arbor, MI, USA*

[*]Corresponding author.

E-mail: lgarmire@med.umich.edu

4 figures
2 tables

# Abstract

Recently developed technologies to generate single-cell genomic data have made a revolutionary impact in the field of biology. Multi-omics assays offer even greater opportunities to understand cellular states and biological processes. However, the problem of integrating different -omics data with very different dimensionality and statistical properties remains quite challenging. A growing body of computational tools are being developed for this task, leveraging ideas ranging from machine translation to the theory of networks and representing a new frontier on the interface of biology and data science. Our goal in this review paper is to provide a comprehensive, up-to-date survey of computational techniques for the integration of multi-omics and alignment of multiple modalities of genomics data in the single cell research field.

# Keywords

Single-cell; Multi-omics; Machine learning; Integration

# Introduction

Single-cell sequencing technologies have opened the door to investigating biological processes at an unprecedentedly high resolution. Techniques such as DROP-seq [1] and 10x Genomics assays are capable of measuring single-cell gene expression, or scRNA-seq, in tens of thousands of single cells simultaneously. Measurements of other data modalities are also increasingly available. For example, single-cell ATAC-seq (scATAC-seq) assesses chromatin accessibility, and single-cell bisulfite sequencing captures DNA methylation, all from single cells. However, many of such techniques are designed to measure a single modality and do not lend themselves to multi-omics measurements. The way to combine information from such measurements is then to assay different -omics from different subsets of the same samples. By assuming that cells assayed by different techniques share similar properties, one can then use alignment methods to computationally aggregate similar cells across different omics assays and draw consensus biological inference.

Recently, however, a number of experimental techniques capable of assaying multiple modalities simultaneously from the same set of single cells have been developed. CITE-seq [2] and REAP-seq [3] measure proteins and gene expression. SNARE-seq [3,4], SHARE-seq [5] and sci-CAR [6] measure gene expression and chromatin accessibility, while scGEM [7] measures gene expression and DNA methylation. For triple-omics data generation, scNMT [8] measures gene expression, chromatin accessibility and DNA methylation, and scTrio-seq [7,9] captures SNPs, gene expression and DNA methylation simultaneously. Integrative analysis of such data obtained from the same cells remains a challenging computational task due to a combination of reasons, such as the noise and sparsity in the assays, and different statistical distributions for different modalities. For clarity, we distinguish between integration methods that combine multiple -omics data from the set of the same single cells (Section I), from alignment methods designed to work with multi-modal data coming from the same tissue but different cells (Section II). The difference in their approaches is shown in Figure. 1.

The application of data fusion algorithms for multi-omics sequencing data predates the single-cell technologies; bulk-level data have been integrated using a variety of computational tools as reviewed in [10]. In this review, we aim to give a comprehensive, up-to-date summary of existing computational tools of multi-omics data integration and alignment in the single-cell field, for researchers in the field of computational biology. For more general surveys, the readers are encouraged to check other single-cell multi-omics reviews [11],[12],[13],[14],[15],[16].

# Integration methods handling multi-omics data generated from the same single cells

The integration methods for multi-modal data assayed from the same set of single cells can be broadly categorized into at least three main types by methodology: mathematical matrix factorization methods, AI (eg. neural-network) based methods and network-based methods. The scheme of these methods is illustrated in Figure 2. Additional less diversified approaches include a Bayesian statistical method and a metric learning method. The list of the currently implemented methods is summarized in Table 1.

## Matrix Factorization based methods

Matrix factorization methods aim to describe each cell as the product between a vector that describes each -omics element (genes, epigenetic loci, proteins, etc.) and a vector of reduced and common features ("factors") capturing its basic properties (Figure 1A). Mathematically, if we represent each -omics as matrix $X_{i\,(i=1,2..)}$ then matrix factorization decomposes it as the product of a shared matrix H across all omics data types, and -omics specific matrix $W_{i\,(i=1,2..)}$, together with random noise $\varepsilon_{i\,(i=1,2,...)}$ as

$$X_1 = W_1 H + \varepsilon_1,\ X_2 = W_2 H + \varepsilon_2, \cdots, X_i = W_i H + \varepsilon_i$$

Such methods are simple and easily interpretable since the cell and -omics factors both carry clearly discernible biological meaning, but may lack the ability to capture nonlinear effects. We describe the variations in this type of methods below:

**MOFA+** [17] is a sequel to the MOFA (Multi-Omics Factor Analysis) [18]. Both studies perform factor analysis, equipped with sparsity-inducing Bayesian elements including Automatic Relevance Determination [19]. MOFA+ integrates data over both views (corresponding to different modalities) and groups (corresponding to different experimental conditions). The model scales easily to large datasets. MOFA+ was applied to integrate gene expression, chromatin accessibility and DNA methylation data assayed using scNMT from mouse embryos, as well as to integrate several datasets over different experimental conditions rather than different -omics. After performing factor analysis on the mouse dataset, the most relevant factors are related to biological processes shaping embryo development. MOFA+ provides an elegant and successful general framework for integration, which could potentially be superseded in specific cases by more specialized models designed for integrating specific -omics layers.

**scAI** ("single-cell aggregation and inference") [20] features a twist on matrix factorization and is designed specifically for integration of epigenetic (chromatin accessibility, DNA methylation) and transcriptomic data. It addresses the sparsity of epigenetic data by aggregating (averaging) such data between similar cells. This requires a notion of cell-cell similarity which is learned as a part of the model, rather than being postulated prior to the integration. Their model solves the following optimization problem

$$\min_{W_1,W_2,H,Z} \alpha ||X_1 - W_1 H||_F^2 + ||X_2(Z \cdot R) - W_2 H||_F^2 + \lambda ||Z - H^T H||_F^2 + \gamma \sum_j ||H_{\cdot j}||_1^2$$

where $X_1$ represents the transcriptomic data, $X_2$ the epigenomic data, H are the common (cell-specific) factors, $W_1$, $W_2$ are the assay-specific factors, Z is the cell-cell similarity matrix, and entries of $R$ are Bernoulli-distributed random variables. The twist on the usual matrix factorization is made by factoring aggregated epigenetic data $X_2$ (Z · R), rather than directly factoring the epigenetic data $X_2$. After the learning is complete, the matrix of cell factors is used to cluster the cells and the importance of genes and epigenetic marks is ranked using the magnitude of the values in loading matrices. In order to jointly visualize different factors, scAI implements a novel VscAI algorithm utilizing Sammon mappings [21]. The relationships between epigenetics and gene expression can be explored using correlation analysis and nonnegative least square regression. The model was tested on simulations using MOSim [22], and several real world datasets, and performed better than the earlier MOFA version, in terms of identifying natural clusters and condensing epigenetic data into meaningful factors.

## Neural Network based methods

While neural networks are generally well-suited for supervised tasks, a class of neural networks called autoencoders is commonly used for unsupervised learning, such as the multi-omics integration problem in single cells. Deep autoencoders perform nonlinear dimensionality reduction by squeezing the input through a lower-dimensional hidden layer ("bottle neck") and attempting to reconstruct the original input as the output of the neural network (Figure 2B). They consist of two parts: the "encoder" network performing the dimensionality reduction and the "decoder" network reconstructing based on the dimensionally reduced data. In principle, autoencoders generalize the principal component analysis by allowing for nonlinear transformations. Many variations of autoencoder models exist, and among them variational autoencoders have proven useful for analyzing single-cell data. Rather than directly encoding the data in a dimensionally reduced ("latent") space, variational autoencoders sample from a probability distribution (usually Gaussian) in the latent space, and use the encoder network to produce the parameters of this distribution. As such, they combine deep learning and Bayesian inference to produce generative models, which not only dimensionally reduce the original data but also produce realistic synthetic data points. Below we review the methods using certain variations of the autoencoder architecture to integrate single-cell multi-omics data.

**scMVAE** ("Single Cell Multimodal Variational Autoencoder") [23] was designed to integrate transcriptomic and chromatin accessibility data, using a version of a variational autoencoder. The key question in multi-omics integration is how to encode the multi-omics data into a single latent space representation. In the case of scMVAE, a combination of 3 different methods was used for this task, including a neural network acting on the concatenated input data, neural networks encoding transcriptomic and chromatin accessibility data separately prior to merging, and a "Product of Experts" technique for combining different representations [24]. At the same time, cell-specific scales used to normalize expression across cells are learned (called "library factors"). The input data are reconstructed by processing the latent representations via decoder neural networks, which calculate the probabilities of gene dropouts and predict the expression of measured genes modelled as a negative binomial distribution.

This model incorporates the task of constructing shared representations of the multi-modal data with clustering. Namely, one of the latent variables is constructed to correspond to the clustering label c. Furthermore, the model incorporates tools to deal with tasks such as data imputation, and can be used for studying the association between epigenetics and gene expression. scMVAE was applied to integrate two real datasets assaying mRNA and chromatin accessibility using SNARE - seq method, as well as simulated data generated by "Splatter" [25]. It takes into account the known relationships between appropriately located transcription factors and gene expression, and uses them to test the imputed (denoised) data. According to the authors, scMVAE performed better than MOFA in terms of clustering and enhancing the consistency between different -omics layers on several real and simulated datasets.

**DCCA,** denoting "<u>D</u>eep <u>c</u>ross-omics <u>c</u>ycle <u>a</u>ttention model", is another method in this category for joint analysis of single-cell multi-omics data [26]. It uses variational autoencoders to integrate multi-omics data, and builds on the scMVAE algorithm described above. However, DCCA diverges from scMVAE in one important aspect: DCCA uses separate but coupled autoencoders to dimensionally reduce different -omics layers, while scMVAE constructs a shared dimensionally reduced representation of transcriptomic and epigenetic data. This strategy is inspired by the theory of machine translation, notably the so-called "attention transfer"; in this case, the "teacher network" working with the scRNA-seq data guides the learning of the "student network" working with scATAC-seq data. Their model compares favorably to scAI and MOFA+ on metrics such as clustering accuracy, denoising quality and consistency between different -omics.

**totalVI** [27] combines Bayesian inference and a neural network to create a generative model for data integration. It was created to handle gene expression and protein data. Joint latent space representations are learned via an encoder network and used to reconstruct the original data while accounting for the difference between the original data modalities. The model generates latent representations capturing both -omics, and at the same time models experimental conditions through an additional set of latent variables. The gene expression data are sampled from a negative binomial distribution, and the parameters are obtained as outputs of a decoder neural network. The protein data are sampled from a mixture model with two negative binomial distributions simulating the experimental background and the actual signal respectively. The model was applied to two datasets containing transcriptomic and proteomic measurements, and generated shared representations of cells with interpretable components.

**LIBRA** [28] uses an autoencoder-like neural network to "translate" between different omics. Motivated by "split-brain autoencoder"[29], and "machine translation" approach, the model consists of two separate neural networks. The first network takes as input elements of the first dataset and aims to reconstruct a corresponding element of the second dataset. The second network performs an inverse task. Taken together, the bottlenecks of two networks aim to convert the two datasets into the same latent space. This method is quite general and can be applied to various pairs of -omics data. It produced clusters of similar quality compared to Seurat v4.

**BABEL** [30] also uses autoencoder-like neural networks to translate between gene expression (modeled by Negative Binomial distribution) and binarized chromatin accessibility data. There are two encoder and two decoder neural networks, each encoder/decoder handles one data type of gene expression or chromatin accessibility. As a result, four combinations between encoders and decoders are formed, and the loss function is optimized to minimize reconstruction error for four combinations of encoders and decoders. In this approach, the two encoders are prone to produce similar representations, as the encoded gene accessibility is decoded as chromatin accessibility and vice versa.

BABEL provides a promising generic framework to multi-omics inference at a single-cell level from single-omics data, by using the model that was previously trained on multi-omics data sequenced from the same single cells. The modular nature of BABEL provides additional flexibility, as the model can be extended to work with additional modalities when the corresponding data becomes available. Despite the potential for generalization, one should be cautioned that if the training is conducted on cell types that are very different, the transfer learning using BABEL is not very successful.

**DeepMAPS** [31] integrates different data modalities by a graph transformer neural network architecture for interpretable representation learning. The data is represented using a heterogenous graph in which some of the nodes represent cells and others represent genes. An autoencoder-like graph neural network architecture is used for representation learning, with an attention mechanism. The attention mechanism learns the weights by the contribution of the neighbors to the node of interest. This not only achieves better performance, but also enhances the interpretability to identify genes most relevant to cell state differences. DeepMAPS method learns relevant gene-gene interaction networks and cell-cell similarities, which can be used for downstream steps such as clustering to infer novel cell types. It compared favorably on clustering, compared to state-of-the art techniques such as MOFA+ and totalVI.

## Network-based methods

Network-based methods represent the relationships between different cells using a weighted graph, where cells serve as nodes (Figure 2C). Integration is then accomplished by manipulating such graph representation. This approach emphasizes the neighborhood structure and sometimes pools the information between neighbors, leading to additional robustness against the noise. Below are the currently available methods.

**citeFUSE** [32] integrates transcriptomic and proteomic CITE-seq data using network fusion of similarity graphs corresponding to different modalities. This idea traces back to computer science work [33] on fusing multi-view networks through cross-diffusion, and to the follow-up SNF method [34] that was used to integrate bulk level multi-omics data. The algorithm adjusts the graph connectivities by a process of diffusion, which allows for the distance information to be aggregated between neighbors. Namely, the algorithm consists of two iterative steps: separate diffusion on different -omics layers and fusion across the -omics layers. It results in a fused consensus matrix of distances between cells, borrowing information from multiple -omics. citeFUSE used spectral clustering to identify cell types, and showed an improvement over single-modality based clusters. Additional benefits of the method include inference of ligand-receptor interactions and a novel tool for doublet detection.

**Joint Diffusion** [35] constructs graph representations of different -omics and then performs a joint diffusion process on the two graphs in order to denoise and integrate the data. This approach builds upon MAGIC [36], a method for denoising scRNA-seq data, and generalizes it to multi-modal data. Diffusion can be conceptualized as a random walk process. In a graph diffusion algorithm, random walking on the graph can help discover the intrinsic structure of the data hidden behind the noise. In Joint Diffusion random walks are performed while allowing for transitions from one graph to another. A key idea in this work is to quantify the amount of noise in different datasets, through a spectral entropy of the corresponding graphs, and adjust the time one spends on different graphs in accordance with their relative levels of noise. In this way, the transcriptomic and epigenetic data will not be weighted equally, as the transcriptomic data is generally of better quality. This method excels at denoising and visualizations, and was shown to present an improved clustering performance compared to single-modality clustering and the one based on a more naive alternating diffusion process.

**Seurat v4** [37] aims to represent the data as a WNN (weighted nearest neighbor) graph in which cells that are similar according to the consensus of both modalities are connected. In the process of constructing a WNN graph, a set of cell-specific weights dictating the relative importance of different -omics data is learned. Such weights often carry important biological meaning. Specifically, Seurat v4 pipeline has the following steps: first, data corresponding to different -omics are dimensionally reduced using PCA to the same number of dimensions. Then, kNN (k nearest neighbor) graphs corresponding to different -omics are constructed. In a kNN graph, each datapoint (a node of this graph) is connected to $k$ nearest neighboring nodes. Cell-specific coefficients determining the relative importance of different -omics are then learned by considering the accuracy of inter-modality and cross-modality predictions by nearest neighbor graphs. Lastly, a linear combination of data from different omics is done, using the coefficients learned in the previous step. The nearest neighbors with respect to those linear combinations are then connected to build the WNN graph. Seurat v4 was applied to a CITE-seq based transcriptomic and proteomic dataset, and several other datasets involving mRNA, proteins and chromatin accessibility. The authors compared this method with MOFA+ and totalVI, using correlations (Pearson and Spearman) between the data corresponding to a cell and the average of its nearest latent space neighbors, and claimed that it performed better than MOFA+ or totalVI.

## Other Models

**BREMSC** [38] is a Bayesian mixture method. It integrates single-cell gene expression and protein data by modeling them as a mixture of probability distributions that share the same underlying set of parameters. The model is useful for performing joint clustering, where confidence in cluster assignments can be quantified via posterior probabilities. It performed favorably compared to single-omics clustering methods. While the MCMC procedure used to train the model can be computationally intensive, the model provides an effective way of integration by accounting the differences between the two -omics layers using probability distributions.

**SCHEMA** [39] is a different metric learning approach that aims to construct a notion of distances on the space of samples, taking into account different -omics data. One of the -omics (usually, scRNA-seq) is considered the primary base for distance, additional omics are then used to modify this distance. This is formulated as optimization of the quadratic function using quadratic programming. The scRNA-seq and scATAC-seq data can thus be integrated, yielding downstream insights into cell developmental trajectories. This method showed a better clustering performance than those based on clustering different modalities separately or integrating them using canonical correlation analysis. It is a useful method for asymmetrically integrating data modalities of different qualities, such as the case of scRNA-seq and scATAC-seq data.

# Alignment methods handling multiple genomics data generated from different single cells of the same tissue

Compared to multi-omics data, it is experimentally much easier to obtain multiple modalities of data where each modality is obtained from similar but different cells of the same tissue. The task to harmonize these data is called alignment (Figure 1). The body of literature applying machine learning and statistical methods to this task is rich, including manifold learning, neural-network based methods, and Bayesian methods, as summarized in Table 2 and depicted in Figure 3. Note that some of the methods developed for batch-correct different scRNA-seq datasets, could in principle be repurposed for single-cell multiple omics alignment; we refer readers to previous benchmark studies [40].

## Bayesian Methods

**Clonealign** [41] integrates single-cell RNA and DNA sequencing data from heterogeneous populations by assigning cells measured by RNA-seq to clones derived from DNA-seq data. Clonealign is based on a Bayesian latent variable model, where a categorical variable is used to specify cell assignment. The model maps the copy number of a gene to its expression value by introducing a copy number dosage effect on the gene expression. The model is also flexible enough to allow for additional covariates such as batch effects or biological information that can be inferred from the gene expression (cell cycle, etc.). In addition to simulation studies that demonstrated robustness, Clonealign was also applied on real cancer datasets to discover novel clone-specific dysregulated biological pathways.

**MUSIC** [42] is an unsupervised topic modeling method for integrative analysis of single-cell RNA data and pooled CRISPR screening data [43]. The model links the gene expression profile of the cells and specific biological function by delineating perturbation effects,, allowing for better understanding of perturbation functions in single cell CRISPR data. In the perturbation effect prioritizing step, MUSIC utilizes the output from the topic model and estimates individual gene perturbation effects on cell phenotypes. It takes three different schemes in modeling combined single-cell and CRISPR data: an overall perturbation effect which represents the gene perturbation effect, a topic model which specifies the function of perturbation effectsway, and with respect to relationships between different perturbation effects. MUSIC was applied to 14 real single-cell CRISPR screening datasets and accurately quantified and prioritized the individual gene perturbation effect on cell phenotypes, with tolerance for substantial noise.

## Manifold Alignment Methods

Manifold alignment methods aim to infer a lower-dimensional structure within multiple complex datasets (Figure 3B). Once this is done, points can be matched across the datasets. This is a very broad class of algorithms, and we here review several representative ones based on distinct ideas, such as the use of pseudotime trajectories, Kernel methods and distance-based matching of cells. The distance-based matching (Figure 4) is a general idea containing several different realizations, such as **UNION-Com** [44], **SCOT** [45] and **Pamona** [46], which are reviewed below, among other methods.

**MATCHER** [47] is the first manifold alignment technique to align different forms of single-cell data. Their approach builds on trajectory inference [48]. It constructs pseudotime trajectories corresponding to cellular processes for each omic first, and then aligns them between different -omics. Pseudotime trajectory models the corresponding cellular process as a Gaussian process and infers the latent variable corresponding to pseudotime. This results in a set of curves capturing the biological processes, one for each -omics layer. Such curves are then projected onto a reference line so that different cells can be matched across -omics. The model makes a strong assumption that there is only one common biological process to be modeled.

**MMD-MA** [49], or Maximum Mean Discrepancy - Manifold Alignment, is a completely unsupervised method. The alignment is performed by matching low-dimensional representations of different -omics, constructed through a kernel-based technique that minimizes the MMD (Maximum Mean Discrepancy) [50] between the two datasets. Additionally, the representations are constructed by taking into account the distortion of the distances in the original data while keeping the transformation as simple as possible. The model was evaluated on data containing gene expression and methylation values from the same single cells; the known cell correspondence information was hidden and MMD-MA was able to successfully reconstruct this information.

**UNION-Com** [44] performs unsupervised alignment of different -omics datasets by matching the structure of the datasets. The idea is that, if different -omics layers indeed correspond to similar samples of cells, then the distance matrices of any two -omics layers will become very similar after rearranging the cell indices. A matching matrix connecting points across datasets is learned by optimizing the similarity of distance matrices after cell permutation. This approach of matching is an extension of GUMA ("Generalized Unsupervised Manifold Alignment") [51] with newly allowed soft matchings. Subsequently, this method performs a version of t-SNE [52] adopted for multi-modal data represented in the same latent space. This approach takes the overall structure of all datasets into account while matching the cells, without the requirement of identical distributions of different modalities. UNION-Com compared favorably with Seurat v3 and MMD-MA when evaluated on the quality of labels transferred between gene expression, methylation and chromatin accessibility data.

**SCOT** [45] is similar to UNION-Com in terms of distance comparison across the -omics layers. However, it is formulated as a different optimization problem per the theory of optimal transport. It starts by considering k-nearest neighbor graphs in different -omics layers and uses those to compute distances between cells from different -omics, like UNION-Com. The soft matchings are applied here as well, with points matched probabilistically across datasets. Unlike UNION-Com, such matchings are obtained by considering a version of optimal transport given by the Gromov-Wasserstein distance, which generalizes the "earth-mover" Wasserstein distance to optimal transport between different spaces [53]. SCOT compared favorably to MMD-MA and UNION-Com on several real and simulated datasets containing transcriptomic and epigenetic (DNAme or chromatin accessibility) data. The model contains only two hyperparameters, making it particularly simple to tune.

**Pamona** [46] uses a similar approach to SCOT, but with a modification of optimal transport based on Partial Gromov-Wasserstein distance [54], which accounts for data points that do not have appropriate matches across datasets. By doing so, the authors can allow for possible imperfect alignment between the datasets, tolerating cell types present in one dataset only. After the alignment is found, the data corresponding to different modalities is projected down to a dimensionally reduced space using Laplacian Eigenmaps [55]. Benchmarked on several datasets containing transcriptomic and epigenetic data, their model outperformed SCOT, MMD-MA and Seurat v3.

## Neural Network-Based Methods

Neural networks, including autoencoders and generative adversarial networks (GAN), have been used for the unsupervised task of the alignment of -omics datasets. Autoencoders have been described earlier. GANs typically consist of two parts: the generator network and the discriminator network. While the generator tries to produce outputs of a form resembling a certain target dataset, the discriminator learns the difference between the generator's outputs and the elements of the target dataset. In this section, we summarize the relevant neural network methods below.

**SCIM** [56] builds on a multi-domain translation approach [57] to integrate multi-omics data in an unsupervised fashion. It uses a separate variational autoencoder for each modality in order to map the data onto reduced latent space representations. Such representations are then aligned to have a similar structure, by using a discriminator network in addition to autoencoders which learns to distinguish between the latent space representations of different -omics. The two autoencoders and the discriminator network are trained simultaneously, resulting in the two latent spaces being maximally alike. Once both datasets are encoded into approximately corresponding representations, the points with similar latent representations are matched across the datasets. This model was tested on simulations from PROSSTT ("Probabilistic Simulation of Single-Cell RNA-seq Tree-Like Topologies") [58] as well as datasets containing gene expression and proteins, and performed favorably to MATCHER when applied to simulated data exhibiting a complex cellular differentiation process.

**MULTIGRATE** [59] uses a multi-modal variational autoencoder structure to project multi-omics data onto a shared latent space. While somewhat similar to the scMVAE model [23], this framework brings additional flexibility and can be used for integration of the paired and unpaired single-cell data. Furthermore, this model can integrate data from a multi-omics assay such as CITE-seq with data from a single-omics assay such as scRNA-seq. Data corresponding to different -omics are first passed through separate neural networks, before being combined by the Product of Experts technique [24] to form the latent distribution. The decoder networks then aim to reconstruct all of the -omics from this unified representation. To better align cells, Maximum Mean Discrepancy is added to the loss function, penalizing the misalignment between the point clouds belonging to different assays. Their model was used for the creation of multi-modal atlases, and mapping a COVID-19 single-cell dataset onto a multi-modal reference.

**MAGAN** [60] utilizes generative adversarial networks (GANs) to align data from different domains. MAGAN uses two tied GANs to translate between the -omics layers, while tying their parameters and requiring that their combination maps any point onto itself. Namely, if the first generator maps data point A to data point B, then the second generator should map B back to A. It is conceptually very similar to the CycleGAN [61] model from computer vision, but with a key innovation that allowed it to more efficiently align and integrate single-cell data. The novelty here was noting that while the CycleGAN framework was very good at aligning the datasets in aggregate, it would not necessarily correctly match individual points. This is a particularly important problem for single-cell data. To address this problem, MAGAN is augmented with a correspondence loss measuring the difference between points before and after being mapped by generators. This model was tested on a variety of datasets, ranging from a simulated dataset to MNIST handwritten digits to molecular data. The method was applied to combine transcriptomic and proteomic data in single cells. The model was shown to meaningfully align the datasets even when the correspondence information was not available.

## Other Methods

**CCA (Canonical Correlation Analysis)** based methods reduce the dimensionality of data by selecting for the degrees of freedom that are correlated between the datasets**. Seurat v3** [62] combines CCA with network concepts in order to align and integrate single-cell multi-omics data. After performing the CCA, the algorithm identifies anchors between the datasets and scores the quality of those anchors. Anchors are identified by MNNs (mutual nearest neighbors), and their quality is scored by considering the overlap between the neighborhoods of anchors. Similar to Seurat v3, MAESTRO [63] also utilized canonical correlation analysis for the integration of transcriptomic and epigenetic data, and provided a comprehensive analysis pipeline. bindSC [64] also uses canonical correlation analysis to construct shared representations of the data, iteratively optimized using a custom procedure.

**LIGER** [65] performs an iNMF (integrative non-negative matrix factorization) to learn factors explaining the variation within and across datasets. Data such as DNA methylation are first aggregated over genes. Cells corresponding to different datasets are described by separate sets of cell-specific factors. Gene factors consist of two components: one that is shared across datasets and one that is dataset specific; the model aims to make the dataset-specific portion as small as possible. After performing the matrix factorization, the shared factor neighborhood graph is formed, in which cells are connected based on the similarity of their factors, and used for aligning the cells across modalities. Recently, this nonnegative matrix factorization approach has been extended to incorporate the idea of online learning. It iteratively updates the model in real-time, and leads to better scalability and computational efficiency[66].

## Concluding Remarks

The landscape of experimental techniques for -omics sequencing and analyzing the data has grown significantly last few years. Accompanying the thrust of technological advancement, an increasing body of computational methods to handle multi-omics data integration or alignment have been proposed. Geared towards computational biologists and genomics scientists, here we reviewed in-depth and extensively these computational methods by their working principles. Among these methods, AI and machine learning based methods account for the majority, demonstrating the influence in single cell computational biology. Other approaches using matrix factorization and or Bayiean's methods have also been proposed. As demonstrated in a range of methods, the integration of multi-omics data at the single-cell level improves the quality of downstream biological interpretation steps, such as clustering. With the advent of technologies for sequencing multi-omics data from the same single cells, efficient multi-omics integration methods to provide further biological and medical insights at larger scales will be of continued demand.

Meanwhile the rapidly growing number of computational methods pose an urgent need for benchmarking studies on their performances, in order to provide guidelines to choose appropriate methods for specific datasets. Current comparisons are either incomplete, or using a small set of benchmark datasets, with inconsistent metrics in various studies, impeding the selection of appropriate methods for the dataset to analyze. This is made more difficult by the generally unsupervised nature of the integration task, where commonly required ground truths are not known for certain. Moreover, different methods have different prerequisites regarding preprocessing steps, normalization, etc and as a result, careful consideration of these steps and their impacts on the model performances is needed. Oftentimes, the integration methods were developed with one specific application/assay in mind, generalization of these methods with the emergence of new technologies needs to be demonstrated. Fortunately, some benchmarking studies have been conducted in other sub-fields of single cell computational biology for reference, such as those focused on the integration of data from different cells and atlas study [67], cell-type annotation [68], and integration algorithms to spatial transcriptomics [69]. Creating standardized high-quality benchmarking datasets would aid such efforts, as proposed in [70] for scRNA-seq data. Finally, comprehensive and flexible benchmarking pipelines that can accommodate the ever-increasing body of integration methods will be extremely useful, in keeping the field up-to-date on multi-omics integration. One such example is the dynverse [71].

## Competing Interests

The authors declare no competing interests.

## Acknowledgements

This work was supported by R01 LM012373 and LM012907 awarded by NLM, and R01 HD084633 awarded by NICHD to L.X. Garmire.

## Bibliography

[1] Macosko EZ, Basu A, Satija R, Nemesh J, Shekhar K, Goldman M, et al. Highly Parallel Genome-wide Expression Profiling of Individual Cells Using Nanoliter Droplets. Cell 2015;161:1202–14.

[2] Stoeckius M, Hafemeister C, Stephenson W, Houck-Loomis B, Chattopadhyay PK, Swerdlow H, et al. Simultaneous epitope and transcriptome measurement in single cells. Nat Methods 2017;14:865–8.

[3] Peterson VM, Zhang KX, Kumar N, Wong J, Li L, Wilson DC, et al. Multiplexed quantification of proteins and transcripts in single cells. Nat Biotechnol 2017;35:936–9.

[4] Chen S, Lake BB, Zhang K. High-throughput sequencing of the transcriptome and chromatin accessibility in the same cell. Nat Biotechnol 2019;37:1452–7.

[5] Clyde D. SHARE-seq reveals chromatin potential. Nat Rev Genet 2021;22:2.

[6] Cao J, Cusanovich DA, Ramani V, Aghamirzaie D, Pliner HA, Hill AJ, et al. Joint profiling of chromatin accessibility and gene expression in thousands of single cells. Science 2018;361:1380–5.

[7] Cheow LF, Courtois ET, Tan Y, Viswanathan R, Xing Q, Tan RZ, et al. Single-cell multimodal profiling reveals cellular epigenetic heterogeneity. Nat Methods 2016;13:833–6.

[8] Clark SJ, Argelaguet R, Kapourani C-A, Stubbs TM, Lee HJ, Alda-Catalinas C, et al. scNMT-

seq enables joint profiling of chromatin accessibility DNA methylation and transcription in single cells. Nat Commun 2018;9:781.

[9] Bian S, Hou Y, Zhou X, Li X, Yong J, Wang Y, et al. Single-cell multiomics sequencing and analyses of human colorectal cancer. Science 2018;362:1060–3.

[10] Huang S, Chaudhary K, Garmire LX. More Is Better: Recent Progress in Multi-Omics Data Integration Methods. Front Genet 2017;8:84.

[11] Statistical single cell multi-omics integration. Current Opinion in Systems Biology 2018;7:54–9.

[12] Ma A, McDermaid A, Xu J, Chang Y, Ma Q. Integrative Methods and Practical Challenges for Single-Cell Multi-omics. Trends Biotechnol 2020;38:1007–22.

[13] Forcato M, Romano O, Bicciato S. Computational methods for the integrative analysis of single-cell data. Brief Bioinform 2021;22:20–9.

[14] Argelaguet R, Cuomo ASE, Stegle O, Marioni JC. Computational principles and challenges in single-cell data integration. Nat Biotechnol 2021. https://doi.org/10.1038/s41587-021-00895-7.

[15] Adossa N, Khan S, Rytkönen KT, Elo LL. Computational strategies for single-cell multi-omics integration. Comput Struct Biotechnol J 2021;19:2588–96.

[16] Miao Z, Humphreys BD, McMahon AP, Kim J. Multi-omics integration in the age of million single-cell data. Nat Rev Nephrol 2021;17:710–24.

[17] Argelaguet R, Arnol D, Bredikhin D, Deloro Y, Velten B, Marioni JC, et al. MOFA+: a statistical framework for comprehensive integration of multi-modal single-cell data. Genome Biol 2020;21:111.

[18] Argelaguet R, Velten B, Arnol D, Dietrich S, Zenz T, Marioni JC, et al. Multi-Omics Factor Analysis-a framework for unsupervised integration of multi-omics data sets. Mol Syst Biol 2018;14:e8124.

[19] Neal RM. Bayesian Learning for Neural Networks. Springer Science & Business Media; 2012.

[20] Jin S, Zhang L, Nie Q. scAI: an unsupervised approach for the integrative analysis of parallel single-cell transcriptomic and epigenomic profiles. Genome Biol 2020;21:25.

[21] A Nonlinear Mapping for Data Structure Analysis n.d. https://ieeexplore.ieee.org/document/1671271 (accessed March 15, 2021).

[22] Martínez-Mira C, Conesa A, Tarazona S. MOSim: Multi-Omics Simulation in R. Cold Spring Harbor Laboratory 2018:421834. https://doi.org/10.1101/421834.

[23] Zuo C, Chen L. Deep-joint-learning analysis model of single cell transcriptome and open chromatin accessibility data. Brief Bioinform 2020. https://doi.org/10.1093/bib/bbaa287.

[24] Hinton GE. Training products of experts by minimizing contrastive divergence. Neural Comput 2002;14:1771–800.

[25] Zappia L, Phipson B, Oshlack A. Splatter: simulation of single-cell RNA sequencing data. Genome Biol 2017;18:1–15.

[26] Zuo C, Dai H, Chen L. Deep cross-omics cycle attention model for joint analysis of single-cell multi-omics data. Bioinformatics 2021. https://doi.org/10.1093/bioinformatics/btab403.

[27] Gayoso A, Lopez R, Steier Z, Regier J, Streets A, Yosef N. A Joint Model of RNA Expression and Surface Protein Abundance in Single Cells. Cold Spring Harbor Laboratory 2019:791947. https://doi.org/10.1101/791947.

[28] Martinez-de-Morentin X, Khan SA, Lehmann R, Tegner J, Gomez-Cabrero D. Machine Translation between paired Single Cell Multi Omics Data. Cold Spring Harbor Laboratory 2021:2021.01.27.428400. https://doi.org/10.1101/2021.01.27.428400.

[29] Split-Brain Autoencoders: Unsupervised Learning by Cross-Channel Prediction n.d. https://ieeexplore.ieee.org/document/8099559 (accessed March 15, 2021).

[30] Wu KE, Yost KE, Chang HY, Zou J. BABEL enables cross-modality translation between multi-omic profiles at single-cell resolution n.d. https://doi.org/10.1101/2020.11.09.375550.

[31] Ma A, Wang X, Wang C, Li J, Xiao T, Wang J, et al. Biological network inference from single-cell multi-omics data using 1 heterogeneous graph transformer 2 2021.

[32] Kim HJ, Lin Y, Geddes TA, Yang J, Yang P. CiteFuse enables multi-modal analysis of CITE-seq data. Cold Spring Harbor Laboratory 2019:854299. https://doi.org/10.1101/854299.

[33] Unsupervised Metric Fusion Over Multiview Data by Graph Random Walk-Based Cross-View Diffusion n.d. https://ieeexplore.ieee.org/document/7348699 (accessed February 17, 2021).

[34] Wang B, Mezlini AM, Demir F, Fiume M, Tu Z, Brudno M, et al. Similarity network fusion for aggregating data types on a genomic scale. Nat Methods 2014;11:333–7.

[35] Kuchroo M, Godavarthi A, Wolf G, Krishnaswamy S. Multimodal data visualization, denoising and clustering with integrated diffusion. arXiv [csLG] 2021.

[36] van Dijk D, Sharma R, Nainys J, Yim K, Kathail P, Carr AJ, et al. Recovering Gene Interactions from Single-Cell Data Using Data Diffusion. Cell 2018;174:716–29.e27.

[37] Hao Y, Hao S, Andersen-Nissen E, Mauck WM, Zheng S, Butler A, et al. Integrated analysis of multimodal single-cell data. Cold Spring Harbor Laboratory 2020:2020.10.12.335331. https://doi.org/10.1101/2020.10.12.335331.

[38] Wang X, Sun Z, Zhang Y, Xu Z, Xin H, Huang H, et al. BREM-SC: a bayesian random effects mixture model for joint clustering single cell multi-omics data. Nucleic Acids Res 2020;48:5814–24.

[39] Singh R, Hie BL, Narayan A, Berger B. Schema: metric learning enables interpretable synthesis of heterogeneous single-cell modalities. Genome Biol 2021;22:131.

[40] Tran HTN, Ang KS, Chevrier M, Zhang X, Lee NYS, Goh M, et al. A benchmark of batch-effect correction methods for single-cell RNA sequencing data. Genome Biol 2020;21:12.

[41] Campbell KR, Steif A, Laks E, Zahn H, Lai D, McPherson A, et al. clonealign: statistical integration of independent single-cell RNA and DNA sequencing data from human cancers. Genome Biol 2019;20:54.

[42] Duan B, Zhou C, Zhu C, Yu Y, Li G, Zhang S, et al. Model-based understanding of single-cell CRISPR screening. Nat Commun 2019;10:2233.

[43] Blei DM, Lafferty JD. A correlated topic model of Science. Aoas 2007;1:17–35.

[44] Cao K, Bai X, Hong Y, Wan L. Unsupervised topological alignment for single-cell multi-omics integration. Bioinformatics 2020;36:i48–56.

[45] Demetci P, Santorella R, Sandstede B, Noble WS, Singh R. Gromov-Wasserstein optimal transport to align single-cell multi-omics data. Cold Spring Harbor Laboratory 2020:2020.04.28.066787. https://doi.org/10.1101/2020.04.28.066787.

[46] Cao K, Hong Y, Wan L. Manifold alignment for heterogeneous single-cell multi-omics data integration using Pamona. Bioinformatics 2021. https://doi.org/10.1093/bioinformatics/btab594.

[47] Welch JD, Hartemink AJ, Prins JF. MATCHER: manifold alignment reveals correspondence between single cell transcriptome and epigenome dynamics. Genome Biol 2017;18:1–19.

[48] Trapnell C, Cacchiarelli D, Grimsby J, Pokharel P, Li S, Morse M, et al. The dynamics and regulators of cell fate decisions are revealed by pseudotemporal ordering of single cells. Nat Biotechnol 2014;32:381–6.

[49] Liu J, Huang Y, Singh R, Vert J-P, Noble WS. Jointly embedding multiple single-cell omics measurements. Cold Spring Harbor Laboratory 2019:644310. https://doi.org/10.1101/644310.

[50] Gretton A, Borgwardt KM, Rasch MJ, Schölkopf B, Smola A. A kernel two-sample test. J Mach Learn Res 2012;13:723–73.

[51] Cui Z, Chang H, Shan S, Chen X. Generalized Unsupervised Manifold Alignment. Adv Neural Inf Process Syst 2014;27.

[52] Van der Maaten L, Hinton G. Visualizing data using t-SNE. J Mach Learn Res 2008;9.

[53] Mémoli F. Gromov–Wasserstein Distances and the Metric Approach to Object Matching. Found Comut Math 2011;11:417–87.

[54] Chapel L, Alaya MZ, Gasso G. Partial Optimal Tranport with applications on Positive-Unlabeled Learning. Adv Neural Inf Process Syst 2020;33:2903–13.

[55] Belkin M, Niyogi P. Laplacian Eigenmaps for Dimensionality Reduction and Data Representation. Neural Comput 2003;15:1373–96.

[56] Stark SG, Ficek J, Locatello F, Bonilla X, Chevrier S, Singer F, et al. SCIM: universal single-cell matching with unpaired feature sets. Bioinformatics 2020;36:i919–27.

[57] Yang KD, Uhler C. Multi-Domain Translation by Learning Uncoupled Autoencoders 2019.

[58] Papadopoulos N, Gonzalo PR, Söding J. PROSSTT: probabilistic simulation of single-cell RNA-seq data for complex differentiation processes. Bioinformatics 2019;35:3517–9.

[59] Lotfollahi M, Litinetskaya A, Theis F. Multigrate: single-cell multi-omic data integration n.d.

[60] Amodio M, Krishnaswamy S. MAGAN: Aligning Biological Manifolds 2018.

[61] Zhu J-Y, Park T, Isola P, Efros AA. Unpaired image-to-image translation using cycle-consistent adversarial networks. Proceedings of the IEEE international conference on computer vision, 2017, p. 2223–32.

[62] Stuart T, Butler A, Hoffman P, Hafemeister C, Papalexi E, Mauck WM, et al. Comprehensive Integration of Single-Cell Data. Cell 2019;177. https://doi.org/10.1016/j.cell.2019.05.031.

[63] Wang C, Sun D, Huang X, Wan C, Li Z, Han Y, et al. Integrative analyses of single-cell transcriptome and regulome using MAESTRO. Genome Biol 2020;21:198.

[64] Dou J, Liang S, Mohanty V, Cheng X, Kim S, Choi J, et al. Unbiased integration of single cell multi-omics data. bioRxiv 2020:2020.12.11.422014. https://doi.org/10.1101/2020.12.11.422014.

[65] Welch JD, Kozareva V, Ferreira A, Vanderburg C, Martin C, Macosko EZ. Single-Cell Multi-omic Integration Compares and Contrasts Features of Brain Cell Identity. Cell 2019;177:1873–87.e17.

[66] Gao C, Liu J, Kriebel AR, Preissl S, Luo C, Castanon R, et al. Iterative single-cell multi-omic integration using online learning. Nat Biotechnol 2021. https://doi.org/10.1038/s41587-021-00867-x.

[67] Luecken MD, Büttner M, Chaichoompu K, Danese A, Interlandi M, Mueller MF, et al. Benchmarking atlas-level data integration in single-cell genomics. bioRxiv 2020:2020.05.22.111161. https://doi.org/10.1101/2020.05.22.111161.

[68] Huang Q, Liu Y, Du Y, Garmire LX. Evaluation of Cell Type Annotation R Packages on Single-cell RNA-seq Data. Genomics Proteomics Bioinformatics 2021;19:267–81.

[69] Li Y, Stanojevic S, He B, Jing Z, Huang Q, Kang J, et al. Benchmarking Computational Integration Methods for Spatial Transcriptomics Data. bioRxiv 2021:2021.08.27.457741. https://doi.org/10.1101/2021.08.27.457741.

[70] Swechha, Mendonca D, Focsa O, Javier Díaz-Mejía J, Cooper S. scMARK an "MNIST" like benchmark to evaluate and optimize models for unifying scRNA data. bioRxiv 2021:2021.12.08.471773. https://doi.org/10.1101/2021.12.08.471773.

[71] Saelens W, Cannoodt R. :: Dynverse n.d. https://dynverse.org/ (accessed January 1, 2022).

# Tables

**Table 1: Summary of the methods for integrating multi-omics data from the same cells**

| Methodology Category | Method | Data | Algorithm | Ref. |
|---|---|---|---|---|
| Matrix Factorization | MOFA+ | Transcriptomic, Epigenetic | Matrix Factorization with Automatic Relevance Determination | [7] |
| | scAI | Transcriptomic, Epigenetic | Matrix factorization, with custom aggregation of epigenetic data | [10] |
| Neural Network | totalVI | Transcriptomic, Proteomic | Variational autoencoder | [12] |
| | scMVAE | Transcriptomic, Epigenetic | Variational autoencoder | [13] |
| | DCCA | Transcriptomic, Epigenetic | Variational autoencoder | [26] |
| | LIBRA | Transcriptomic, Proteomic, Epigenetic | Split-brain autoencoder | [16] |
| | BABEL | Transcriptomic, Proteomic, Epigenetic | Autoencoder translating between modalities | [30] |
| | DeepMAPS | Transcriptomic, Epigenetic, Proteomic | Graph Neural Network | [31] |
| Network - Based | citeFUSE | Transcriptomic, Proteomic | Similarity network fusion | [18] |
| | Seurat v4 | Transcriptomic, Proteomic | Weighted averaging of nearest neighbor graphs | [20] |
| | Integrated Diffusion | Transcriptomic | Joint Manifold Learning through Integrated Diffusion | [35] |
| Other | BREM-SC | Transcriptomic, Proteomic | Bayesian mixture model | [38] |
| | SCHEMA | Transcriptomic, Epigenetic | Metric Learning | [39] |

**Table 2: Summary of the computational methods for aligning multiple omics data from different single cells**

| Methodology Category | Method | Algorithm | Data | Ref. |
|---|---|---|---|---|
| Manifold Alignment | UNION - Com | Topological Alignment | Transcriptomic, Epigenetic | [44] |
| | MATCHER | Pseudotime Reconstruction and Manifold Alignment | Transcriptomic, Epigenetic | [47] |
| | MMD-MA | Manifold Alignment | Transcriptomic, Epigenetic (DNAme) | [49] |
| | SCOT | Gromov-Wasserstein optimal transport | Transcriptomic, Epigenetic (DNAme, accessibility) | [45] |
| | Pamona | Partial Gromov-Wasserstein optimal transport | Transcriptomic, Epigenetic | [46] |
| Neural Network | MAGAN | Generative Adversarial Network | Transcriptomic, Proteomic | [60] |
| | SCIM | Adversarial autoencoder | Transcriptomic, Proteomic (CyTOF) | [56] |
| | Multigrate | Variational Autoencoder | Transcriptomic, Proteomic | [59] |
| Bayesian | clonealign | Bayesian latent variable model | RNA-seq, DNA | [41] |
| | MUSIC | Topic models | RNA, CRISPR | [50] |
| Other | Seurat v3 | Canonical Correlation Analysis and Mutual Nearest Neighbors analysis | RNA-seq, ATAC-seq | [62] |
| | bindSC | | RNA-seq, ATAC-seq | [64] |
| | MAESTRO | Canonical Correlation Analysis | RNA-seq, ATAC-seq | [63] |
| | LIGER | Matrix factorization | RNA-seq, methylation | [62, 65] |

# Figures and Figure Legends

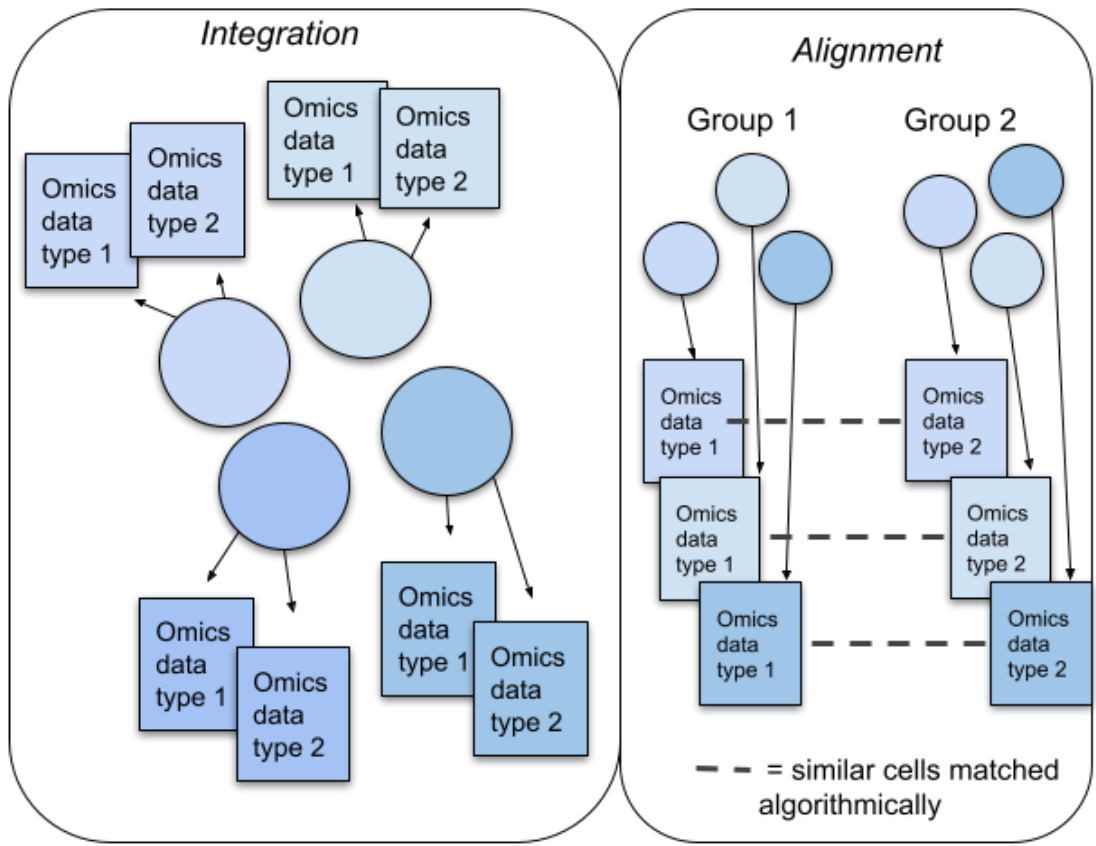

**Figure 1: Multi-omics data can sometimes be sequenced from the same set of single cells (left); at other times, only the data sequenced from the same/similar sample, but different single cells are available (right).**

In the former case, we have the task of integrating the different data modalities (left); in the latter case, we need to first identify similar cells across the samples (right) - this is the computational task of alignment.

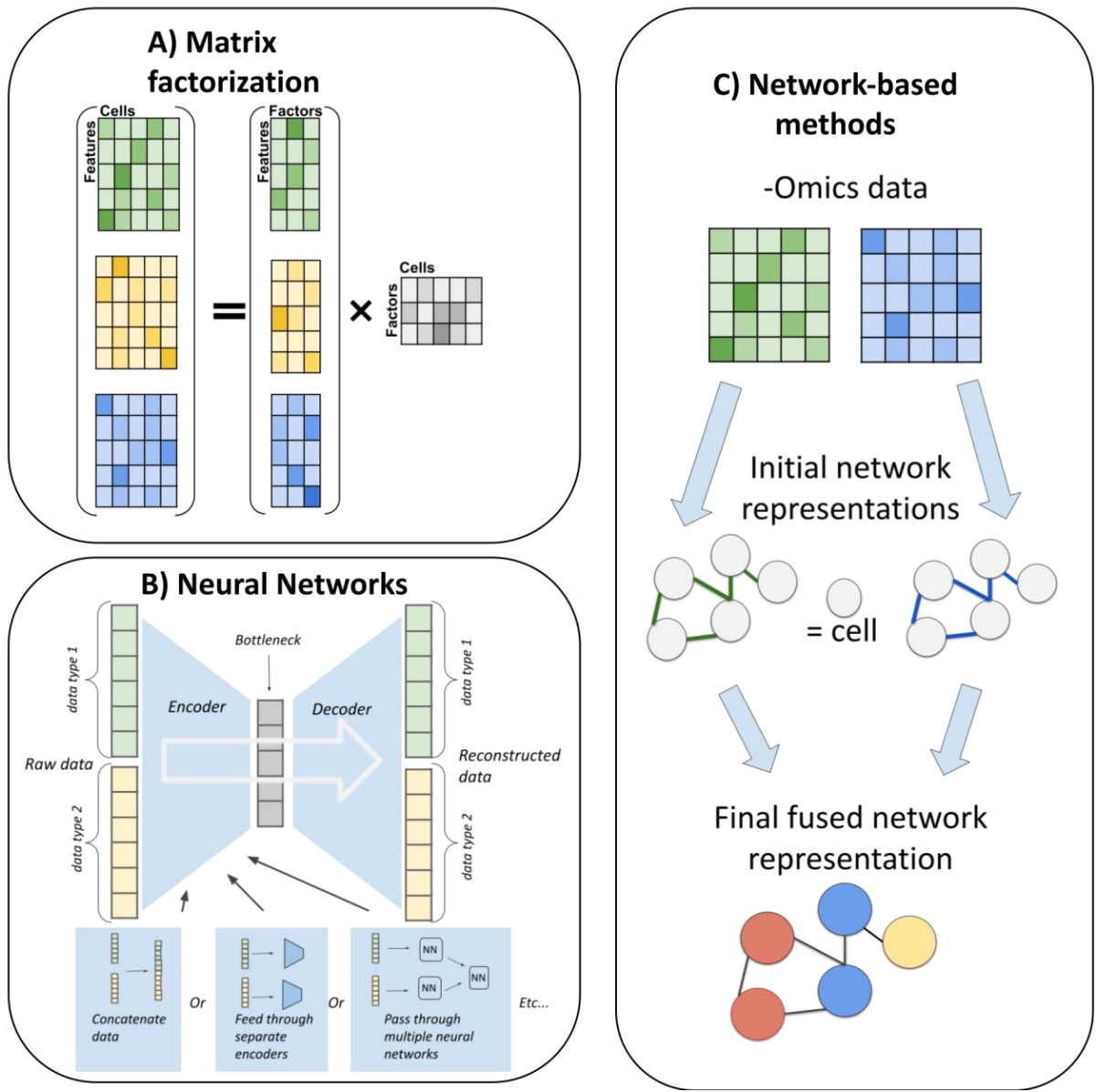

**Figure 2: Illustration of some common integration approaches for single-cell multi-omics**

A) matrix factorization, B) neural networks and C) network-based approaches.

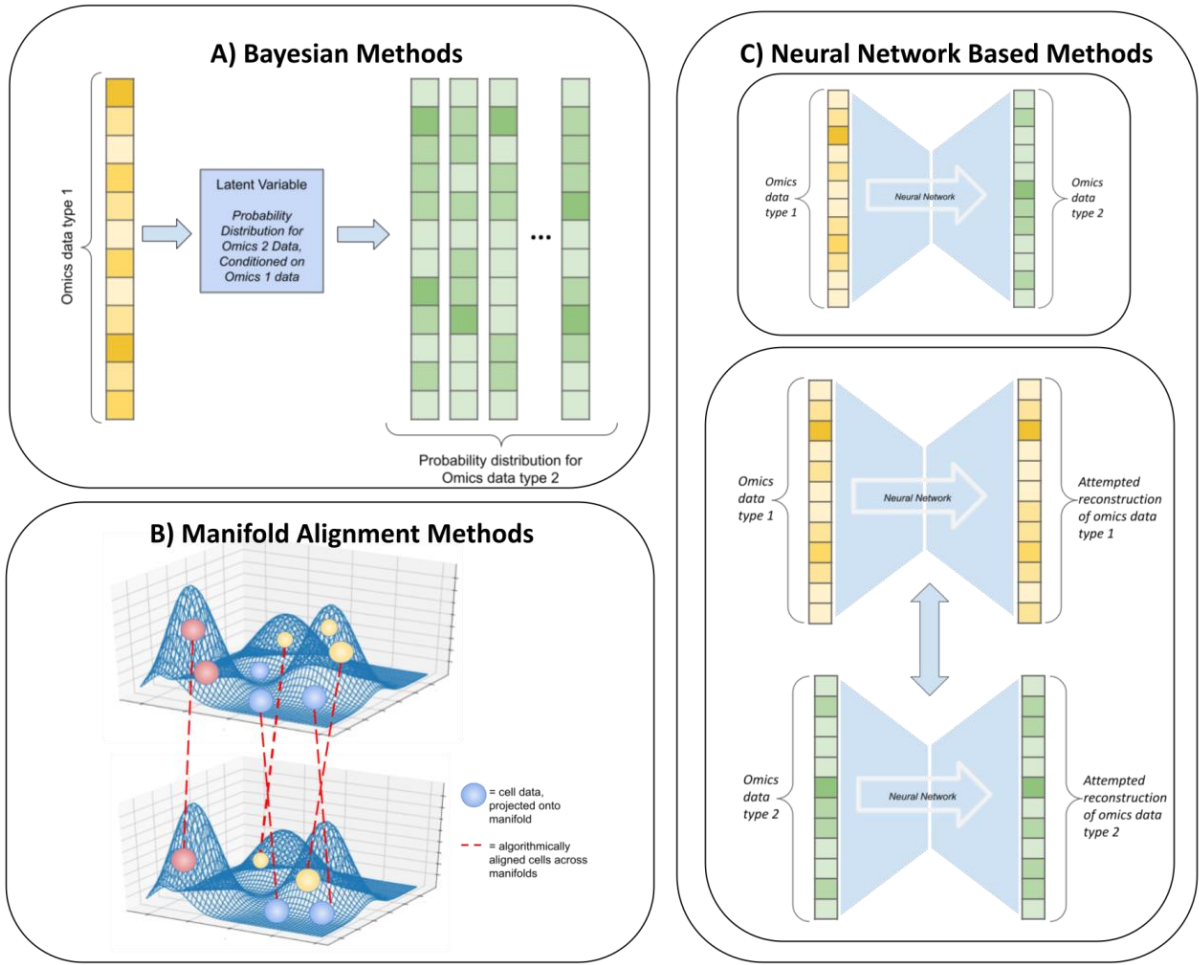

**Figure 3: Illustration of some common approaches for alignment of multi-omics single-cell data**

A) Bayesian methods, B) manifold alignment methods and C) neural network based models.

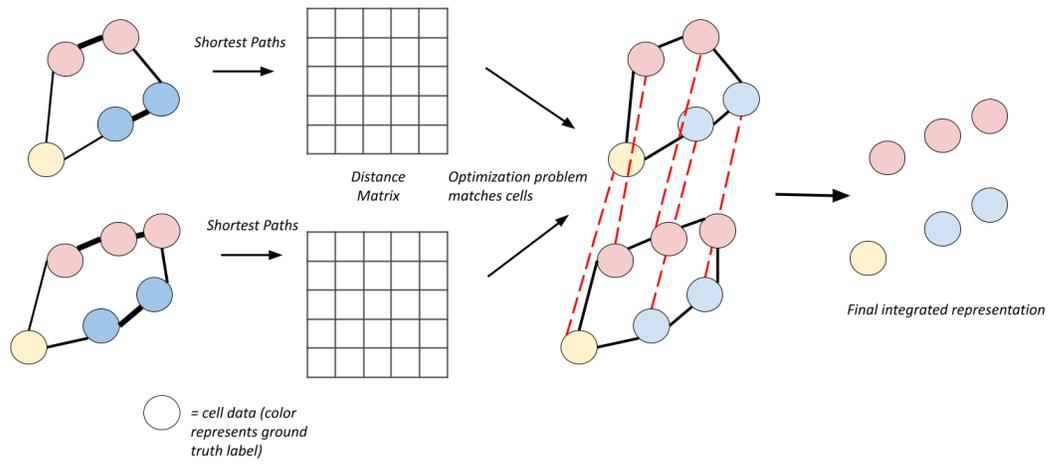

**Figure 4: Summary of the distance-based alignment algorithm**

Cells are represented by nodes in two different graph representations and matched in order to preserve a notion of the cell-cell distance on the graph.